\documentclass[aps,prb,twocolumn,10pt,longbibliography,floatfix,nofootinbib,superscriptaddress]{revtex4-2}

\usepackage{titlesec}
\titleformat{name=\section,numberless}[block]{\normalfont\bfseries}{\thesection}{1em}{\filright}

\usepackage{physics}
\usepackage{float}
\usepackage{bbold}
\usepackage[english]{babel}
\usepackage[utf8]{inputenc}
\usepackage{amssymb}
\usepackage{amsmath}
\usepackage{dsfont}
\usepackage{multirow,tabularx,makecell}
\usepackage[pdftex]{graphicx}
\usepackage{xspace}
\usepackage{dsfont}
\usepackage{bm}
\usepackage{nicefrac}
\usepackage[export]{adjustbox}
\usepackage{makecell}

\usepackage{tabularx} 
\usepackage{makecell} 
\usepackage{graphicx} 
\usepackage{array}    
\newcommand{\xmark}{\text{\sffamily X}}

\usepackage[pdfencoding=auto, psdextra]{hyperref}
\hypersetup{
    colorlinks,%
    citecolor=blue,%
    filecolor=blue,%
    linkcolor=blue,%
    urlcolor=blue
}

\usepackage[usenames, dvipsnames]{xcolor}

\providecommand{\blue}[1]{\textcolor{black}{#1}}

\begin{document}

\title{Circular motion of non-collinear spin textures in Corbino disks: Dynamics of Néel- versus Bloch-type skyrmions and skyrmioniums}

\author{Ismael Ribeiro de Assis}
\email[Correspondence email address: ]{ismael.ribeiro-de-assis@physik.uni-halle.de}
\affiliation{Institut f\"ur Physik, Martin-Luther-Universit\"at Halle-Wittenberg, D-06099 Halle (Saale), Germany}

\author{Ingrid Mertig}
\affiliation{Institut f\"ur Physik, Martin-Luther-Universit\"at Halle-Wittenberg, D-06099 Halle (Saale), Germany}

\author{B{\"o}rge G{\"o}bel}
\email[Correspondence email address: ]{boerge.goebel@physik.uni-halle.de}
\affiliation{Institut f\"ur Physik, Martin-Luther-Universit\"at Halle-Wittenberg, D-06099 Halle (Saale), Germany}

\date{\today}

\begin{abstract}
Magnetic skyrmions are nano-scale magnetic whirls that can be driven by currents via spin torques. They are promising candidates for spintronic devices such as the racetrack memory, where a motion along the uniform current is typically desired. However, \blue{for spin torque nano-oscillators} in Corbino disks, the goal is to achieve a circular motion, perpendicular to the radially applied current. As we show, based on analytical calculations and micromagnetic simulations, Bloch skyrmions engage in a circular motion with frequencies in the MHz range when driven by spin-orbit torques. In contrast, Néel skyrmions get stuck at the edges of the disk. Our analysis reveals that the antagonistic dynamics between Bloch- and Néel-type magnetic textures arise from their helicity. Furthermore, we find that skyrmioniums, which are topologically trivial variations of skyrmions, move even faster and allow an increase in the current density without being pushed toward the edges of the disk. When driven by spin-transfer torques instead, Bloch and Néel skyrmions no longer exhibit different dynamics. Instead, they move along a circular trajectory due to the skyrmion Hall effect caused by their topological charge. Consequently, the topologically trivial skyrmioniums inevitably become trapped at the disk edge in this scenario. To provide a comprehensive understanding, our study also examines currents applied tangentially, further enriching our insights into skyrmion dynamics and appropriate current injection methods for skyrmion-based devices.
\end{abstract}

\maketitle

\section{Introduction}

\begin{figure*}[t!]
  \centering
    \includegraphics[width=\textwidth]{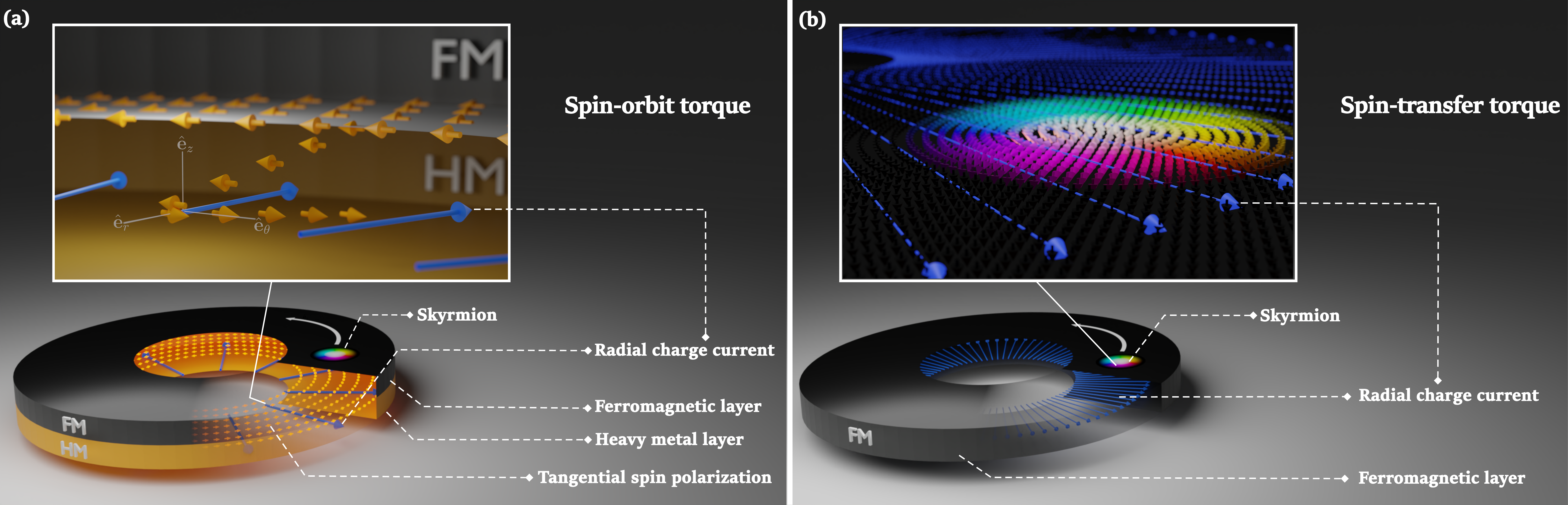}
    \caption{Technologically relevant scenarios of skyrmion motion driven by spin torques in Corbino disk geometries, as discussed in this work. (a) Spin-orbit torque scenario: This panel schematically illustrates the interface between ferromagnetic (FM) and heavy metal (HM) layers. As shown in the inset, a radial charge current (blue arrow) in the HM layer induces a spin current directed towards the FM layer, resulting from the spin Hall effect. Consequently, a tangential spin polarization (orange arrows) arises in the FM layer. The spin current induces a torque, allowing a skyrmion to engage in a circular motion along the disk's edges at MHz frequency, as detailed in Sec.~\ref{section:SOTskyrmion}. (b) Spin transfer torque scenario: This panel schematically displays the charge current (blue arrows) flowing radially through the FM layer and through the skyrmion, as shown in the inset. The radial flow of conduction electrons induces a torque that facilitates the circular motion of skyrmions, as elaborated in Sec.~\ref{subsection:STT}.}
    \label{fig:schematic_scenarios}
\end{figure*}

Magnetic skyrmions are quasi-particles notable for their high stability, compact size, and particle-like behavior, characterized by their unique topology. They can be generated and moved using currents~\cite{kern2022deterministic,gobel2019electrical,jonietz2010spin,jiang2017direct,litzius2017skyrmion} or gradients in the magnetic parameters~\cite{muller2016edge,schaffer2020rotating,yu2016room,de2023skyrmion,tomasello2018chiral,gorshkov2022dmi,zhang2023current} and have promising potential for future data storage and computing applications~\cite{fert2013skyrmions,bobeck1969magnetic,michaelis1975magnetic,parkin2004shiftable,li2017magnetic,de2023biskyrmion,msiska2023audio,azam2018resonate,song2020skyrmion}. Moreover, their topology directly influences their dynamics when driven by spin torques and even different types of skyrmions, such as Néel~\cite{heinze2011spontaneous} and Bloch~\cite{muhlbauer2009skyrmion}, can exhibit distinct propagation directions.

Despite their potential for spintronics, a significant shortcoming in skyrmion-based applications, particularly in racetrack memories, is the skyrmion Hall effect (SkHE) \cite{jiang2017direct,litzius2017skyrmion}
. This effect causes skyrmions to move transversely to an external force, potentially destroying them at the device's edges. Addressing this challenge has led to extensive research to mitigate it. Considered approaches include but are not limited to, torque engineering~\cite{gobel2019overcoming}, nano-patterning~\cite{juge2021helium, kern2022deterministic,ahrens2022skyrmion}, and using non-conventional skyrmions~\cite{gobel2021beyond}, such as biskyrmions~\cite{yu2014biskyrmion,wang2016centrosymmetric}, skyrmioniums~\cite{finazzi2013laser,zheng2017direct,zhang2018real} or antiskyrmions~\cite{nayak2017magnetic}. Nonetheless, alternative device concepts, like a skyrmion-based ratchet~\cite{gobel2021skyrmion} or an artificial neuron~\cite{de2023biskyrmion}, have shown that leveraging this effect can be advantageous.

Another innovative approach involves using alternative geometries, such as nanodisk configurations. These so-called 'Corbino disk' geometries have provided insights into skyrmion creation~\cite{zhang2018manipulation,ponsudana2021confinement,talapatra2018scalable} and dynamics~\cite{kechrakos2023skyrmions,paikaray2021skyrmion,cai2023nontraditional,xia2019current,jiang2018dynamics} and offer a new platform for spintronic devices~\cite{kechrakos2023skyrmions,feng2019skyrmion,jin2020high,liang2020spiking}. In racetrack memories, the current is constrained to a homogeneous direction. In contrast, Corbino disks allow the application of a current that flows from the center of the disk along the radial direction. Skyrmions can undergo circular motion in these structures, with frequency and rotation direction determined by the strength and sign of the applied current. \blue{As a result, skyrmions may play a role in promising applications such as spin torque nano-oscillators (STNOs)~\cite{katine2008device,chen2016spin,georges2008coupling}. These devices have attracted significant research interest due to their versatility, serving as wireless communication devices~\cite{litvinenko2021analog}, magnetic field sensors~\cite{braganca2010nanoscale}, and artificial neurons~\cite{bohnert2023weighted,liang2020spiking}, among other uses. Skyrmion-based STNOs have been proposed in several works~\cite{liang2020spiking,garcia2016skyrmion,chen2019magnetic,shen2019spin}. However,} a comprehensive understanding of the precise conditions facilitating \blue{the rotational} motion for different types of skyrmionic textures remains elusive.


In this paper, we address this issue numerically and analytically by examining the behavior of Néel and Bloch skyrmions and skyrmioniums in a Corbino geometry under various spin torques generated by applied charge currents, as schematically summarized in Fig.~\ref{fig:schematic_scenarios}. First, in Sec.~\ref{section:SOTskyrmion}, we demonstrate that Néel and Bloch skyrmions exhibit opposite behaviors when driven by spin-orbit torques (SOTs) generated by the spin Hall effect in a bilayer geometry. We reveal that their helicity combined with the direction of spin polarization is decisive for the resultant dynamics, determining whether a skyrmion engages in circular motion or halts at the edges. While a Bloch skyrmion engages in a circular motion for a radial current, the Néel skyrmion does so only for a tangential current. Moreover, we obtain an analytical equation characterizing the circular motion for any skyrmion-helicity using the Thiele equation. 

Second, by expanding our analysis beyond conventional skyrmions to topologically trivial skyrmioniums in Sec.~\ref{subsection:skyrmioniums}, we show that Néel- and Bloch-type skyrmioniums follow the same antagonistic behavior when driven by SOTs. However, unlike conventional skyrmions that move along the disk's edges due to the SkHE, skyrmioniums move directly in the direction of the SOT-induced force without needing to interact with the edges due to their vanishing topological charge. We highlight the role of helicity in their dynamics, giving insights into selecting appropriate materials and current injection methods for skyrmion-based devices from circular geometries to rectangular racetrack memories. 

Third, in Sec.~\ref{subsection:STT}, we discuss the dynamics by a current-induced spin transfer torque (STT) in which Néel and Bloch skyrmions exhibit identical dynamics. For this alternative scenario, our analysis reveals that the determining factor for a circular motion is the SkHE caused by the topological charge. Unlike in the SOT scenario, we show that Néel- and Bloch-type skyrmioniums do not echo the dynamics of their standard skyrmion counterparts. We find that, under radial currents, the skyrmion Hall effect—often regarded as detrimental—facilitates circular motion in skyrmions, while skyrmioniums get stuck at the disk edges. These results underscore the nuanced impact of different torques on skyrmion and skyrmionium behavior.

The paper is structured as follows: We first highlight the distinctions between Néel and Bloch skyrmions in Sec.~\ref{subsection:Néel_and_Bloch} and show how their helicity influences their motion under SOT in Sec.~\ref{subsection:thiele_equation}. Subsequently, in Sec.~\ref{subsection:Corbino}, we examine analytically the dynamics of both skyrmions in the Corbino disk geometry under radial and tangential current orientation. We confirm these predictions by micromagnetic simulations in Sec.~\ref{subsection:micromagnetic_simulations}. We generalize the discussion of the antagonistic dynamics in the SOT-driven scenario to skyrmioniums in Sec.~\ref{subsection:skyrmioniums}. Finally, in Sec.~\ref{subsection:STT}, we discuss the congruent dynamics of Bloch- and Néel-type textures when driven by STT with currents along the radial and tangential directions.

\section{Current-induced rotational motion of skyrmions driven by spin-orbit torques}\label{section:SOTskyrmion}

In the first main part of this paper, we analyze the motion of different types of skyrmions by spin-orbit torques. We begin by introducing how Bloch and Néel skyrmions can be distinguished and characterized by their helicity.

\subsection{Néel skyrmions and Bloch skyrmions}
\label{subsection:Néel_and_Bloch}

\begin{figure}[t!]
  \centering
    \includegraphics[width=\columnwidth]{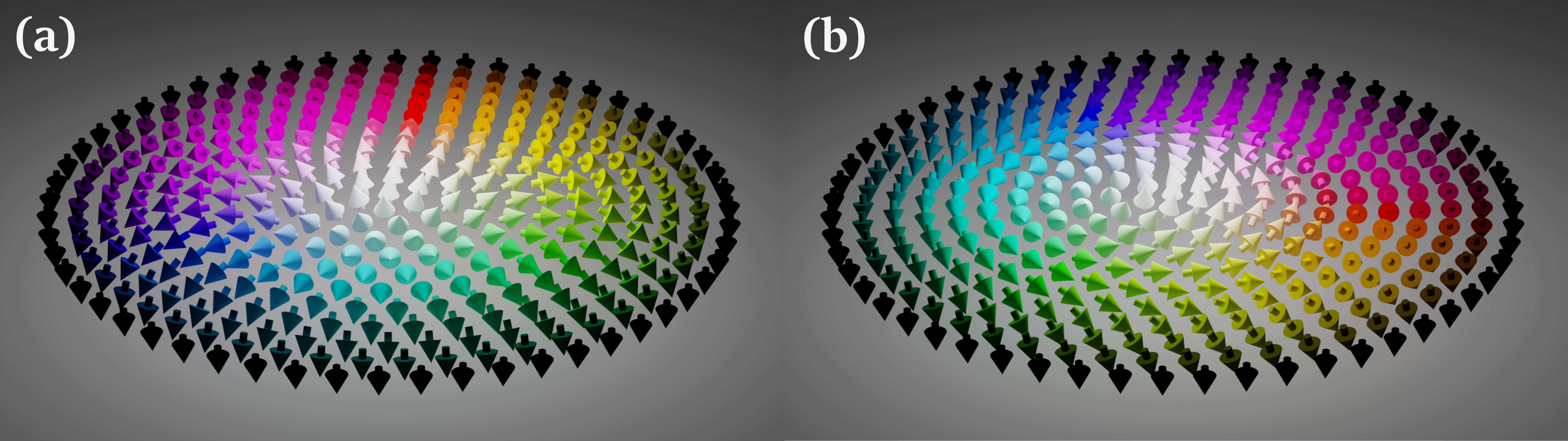}
    \caption{Illustrative comparison of two types of magnetic skyrmions. Schematic representation of (a) Néel skyrmion and (b) Bloch skyrmion. Arrows correspond to the orientation of magnetic moments, which is also encoded by the color. Out-of-plane: Black and white. In-plane: color spectrum depending on the in-plane angle. }
    \label{fig1:skyrmions}
\end{figure}

Skyrmions are non-collinear magnetic textures characterized by a topological charge $N_\mathrm{Sk}$ and a set of geometrical quantities \cite{nagaosa2013topological,gobel2021beyond}: polarity $p$, helicity $\gamma$, and vorticity $m$. 
Skyrmions owe their stability to the topological charge defined via
\begin{equation}
N_\mathrm{Sk} = \frac{1}{4 \pi }\int \bm{m} \cdot (\partial_x \bm{m} \times \partial_y \bm{m} ) \, \mathrm{d}^2 r,    
\end{equation}
or in terms of the geometric quantities, 
\begin{equation}
    N_\mathrm{Sk} = p \cdot m,
\end{equation}
where $m = 0, \pm 1 , \pm 2, ...$ and $p=\pm 1$.

The polarity $p$ specifies the magnetization orientation at the skyrmion core along the z-axis, with $p = \pm 1$, opposite to the ferromagnetic background. The vorticity $m$ indicates the sense of in-plane rotation of the magnetization when tracing magnetic moments in a closed loop around the skyrmion core. The helicity $\gamma$ is a phase difference between the polar angle in the magnetization space $\vartheta$ and the polar angle of the position vector $\theta$. The magnetization in spherical coordinates is $\bm{m}(\bm{r}) = ( \sin \phi \cos(m\theta + \gamma), \sin \phi \sin(m\theta + \gamma), \cos \phi )$ where $\phi(r)$ is the azimuthal angle of the magnetic texture that is determined by the magnetic parameters of the sample.

Among the several types of magnetic skyrmions \cite{gobel2021beyond}, in this work, we focus on two specific skyrmions, namely, Néel and Bloch skyrmions, illustrated in Fig.~\ref{fig1:skyrmions} (a) and (b), respectively. Both non-trivial textures have a vorticity of $m=1$ leading to the same topological charge of $N_\mathrm{Sk} = \pm 1$, depending on the magnetic background $p=\pm 1$. 

The unique dynamics of skyrmions can be significantly influenced by their geometrical characteristics. Given a ferromagnetic background aligned along the negative z-axis and a topological charge $N_\mathrm{Sk} = + 1$, both skyrmions have the same polarity $p = +1$, and vorticity $m = + 1$—their distinction lies at the helicity [Fig.~\ref{fig1:skyrmions}]. Specifically, Néel skyrmions have a helicity $\gamma$ of either 0 or $\pi$, while Bloch skyrmions have a helicity of $\pm \pi/2$, where the sign indicates anti-clockwise and clockwise in-plane rotation, respectively. In this work, we shall consider Néel skyrmions with $\gamma =0$ and Bloch skyrmions with $\gamma = \pi/2$. 

\subsection{Thiele equation}
\label{subsection:thiele_equation}
In the following, we describe in detail the effective equation of motion of a skyrmion, namely, the Thiele equation, given by
\begin{equation}\label{eq:Thiele}
    b \underline{\bm{G}} \times \bm{v} - b \underline{\bm{D}} \alpha \bm{v} - B j \underline{\bm{I}} \bm{s}  = \bm{F}_e,     
\end{equation}
where $b$ and $B$ are constants determined from the sample parameters with $b = M_s d_z / \gamma_e$ and $ B = \hbar \Theta_\mathrm{SH} / 2 e $ . Here, $M_s$ is the saturation magnetization, $d_z$ the thickness of the ferromagnetic sample, $\gamma_e$ is the gyromagnetic ratio, and $\Theta_\mathrm{SH}$ is the spin Hall angle.

The first term is the so-called gyroscopic force, characterized by the gyroscopic vector $\bm{G} = 4 \pi N_\mathrm{Sk} \hat{\bm{e}}_z$ arising from the topological charge. The second term is the dissipative force, quantified by the Gilbert damping $\alpha$. The dissipative tensor 
$D_{ij} = \int ( \partial_{x_i} \bm{m} \cdot \partial_{x_j} \bm{m} ) \, \mathrm{d}^2 r$
only has non-zero $D_{xx}=D_{yy}\equiv D_0$ elements, irrespective of the type of skyrmion, as long as there is no deformation. The third term accounts for the SOT and will be explained below. The fourth term represents a force arising due to the environment in which the skyrmion resides (accounting for edges, defects, other skyrmions, etc). Here, we regard this term as the interaction with the edges; a repulsive force is exerted on the skyrmion as it approaches the edges.

A standard approach for moving skyrmions uses SOTs due to the low currents required \cite{sampaio2013nucleation}. The third term in the Thiele equation quantifies this effect. In this scenario, the ferromagnetic layer (FM) hosting the skyrmion is interfaced with a heavy metal layer (HM). To induce SOTs in the FM layer, a charge current $\bm{j}$ is injected into the HM, giving rise to a spin current $j_s = j \theta_{\text{SH}}$ in the FM along the $z$-direction due to the spin Hall effect. The spin polarization $\bm{s}$ is oriented along $\bm{j} \times \hat{\bm{e}}_z$, i.e., perpendicular to the charge current direction. We proceed by analyzing this term in greater detail.

The torque tensor $I_{ij} = \int ( \partial_{x_i} \bm{m} \times \bm{m} )_j \,\mathrm{d}^2 r$ is directly related to the skyrmion helicity by \blue{\cite{kim2018asymmetric,ritzmann2018trochoidal}}
\begin{align}
\label{eq:torque_tensor_helicity}
\underline{\bm{I}} = I_0 \begin{pmatrix}
\sin{\gamma} & -\cos{\gamma} \\
 \cos{\gamma} & \sin{\gamma}
\end{pmatrix},
\end{align}
where $I_0$ is a constant roughly determined by the skyrmion size.

For Bloch skyrmions, the tensor is diagonal, while for Néel skyrmions, the tensor is off-diagonal. It is worth noting that this term appears accompanied by the spin polarization of the induced spin current in the Thiele equation~\eqref{eq:Thiele}. Consequently, due to the tensor's diagonality, the force caused by the SOT aligns either anti-parallel or parallel with the spin polarization for Bloch skyrmions. In contrast, for Néel skyrmions, the off-diagonal character of the tensor results in a force orthogonal to the spin polarization. As we will discuss next, this nuance is the reason for the distinct dynamics of Néel and Bloch skyrmions.

\subsection{Predicted dynamics in Corbino disks}
\label{subsection:Corbino}

In this section, we investigate the dynamics of Néel and Bloch skyrmions under SOT in Corbino disks. We consider the stabilized skyrmions to be in an FM layer on top of an HM layer, both of which conform to a Corbino disk shape. For instance, a current can be induced in the HM by placing contacts at the inner and outer edges of the disk. In this case, the current flows along the radial direction, thereby inducing a tangential spin polarization due to the spin Hall effect, as schematically illustrated in Fig.~\ref{fig:schematic_scenarios} (a). Additionally, to comprehensively understand the skyrmionic dynamics, we consider an alternative yet less technologically relevant scenario where a tangential charge current leads to a radial spin polarization. In the following, we establish the analytical framework for both scenarios based on the Thiele equation.

Due to the Corbino disk geometry, expressing the Thiele equation in polar coordinates is convenient. Therefore, we rewrite Eq.~\eqref{eq:Thiele} as
\begin{align}
\nonumber
  -4 \pi  N_\mathrm{Sk} b
&\begin{pmatrix}
 - r  \dot{\theta} \\
\dot{r}
\end{pmatrix}
- b D_{0} \alpha 
\begin{pmatrix}
 \dot{r}\\
r  \dot{\theta}
\end{pmatrix}
=
\\
&B j I_0 \begin{pmatrix}
\sin{\gamma} & -\cos{\gamma} \\
\cos{\gamma} & \sin{\gamma}
\end{pmatrix}
\begin{pmatrix}
s_r\\
s_{\theta}
\end{pmatrix} 
+ \begin{pmatrix}
 F_e\\
0
\end{pmatrix}.
\end{align}
%

\begin{figure*}[t!]
  \centering
    \includegraphics[width=\textwidth]{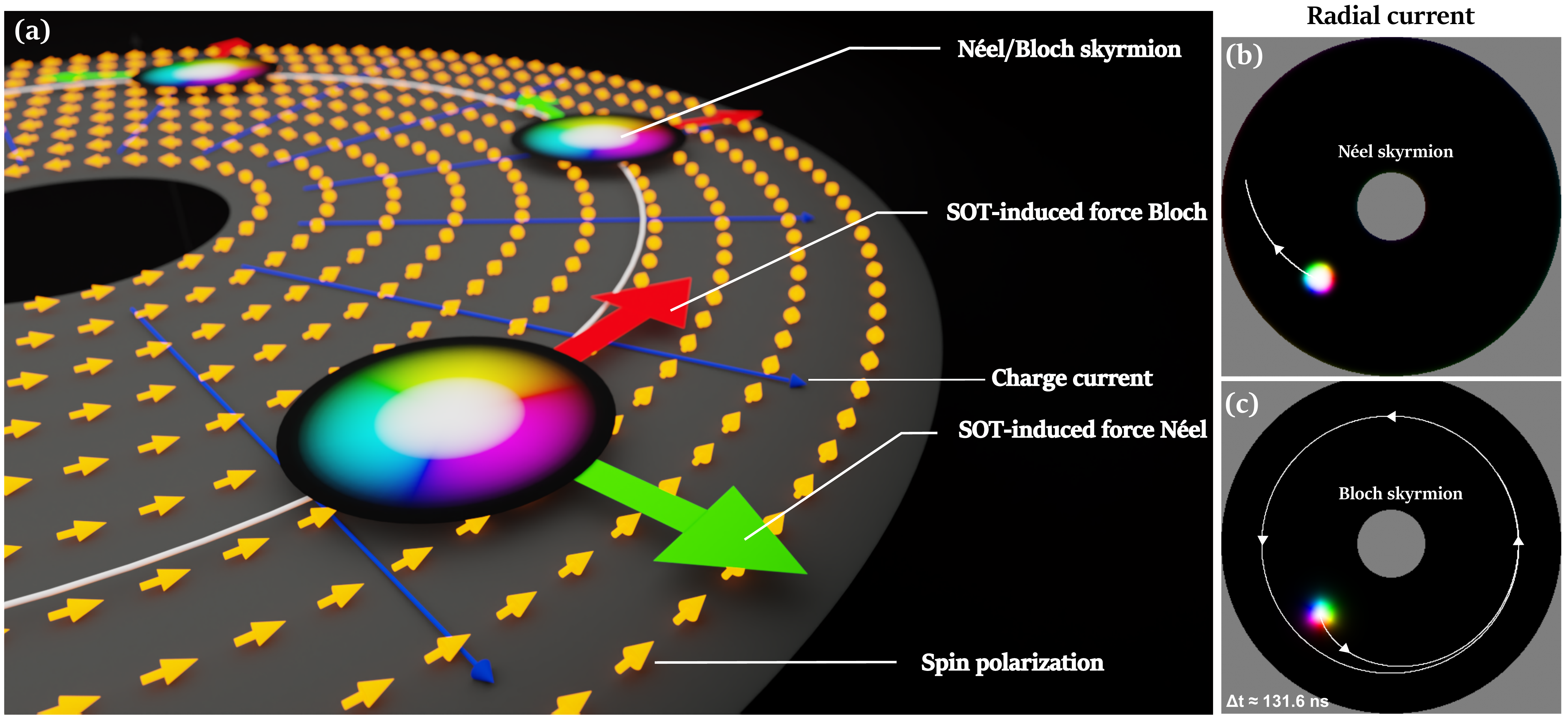}
    \caption{Dynamics of Néel and Bloch skyrmions under radial currents giving rise to a tangential spin polarization. (a) Schematic representation of Néel and Bloch skyrmions subjected to spin currents with tangential polarization. Néel skyrmions are influenced by a radial SOT force (green arrow), while Bloch skyrmions experience a tangential SOT force (red arrow). (b) and (c) show the trajectories of Néel and Bloch skyrmions, respectively, obtained by micromagnetic simulations for 20 ns for (b) and 200 ns for (c). The marker at the lower left corner of subplot (c) indicates the time duration for completing one loop around the disk with $\Delta t \approx 131.6$ ns for the Bloch skyrmion.}
    \label{fig:spin_tangential}
\end{figure*}

Here, $s_r$ and $s_{\theta}$ denote the spin polarization's radial and tangential unit vector components. The edge force $F_e$ acts solely along the radial direction due to the disk's geometry. Solving this system of equations yields the radial velocity $\dot{r}$ 
\begin{align}
\nonumber
    \dot{r} = \Gamma \Big[ -D_0 \alpha F_e &+ I_0 B j \Big (\cos{\gamma }( D_0 \alpha s_{\theta} - 4 \pi  N_\mathrm{Sk} s_r ) \\
    &- \sin{\gamma }( 4 \pi  N_\mathrm{Sk} s_{\theta} + D_0 \alpha s_{r} ) \Big) \Big],
\end{align}
and the angular velocity $\dot{\theta}$
\begin{align}
\nonumber
    \dot{\theta} = \frac{\Gamma}{r} \Big[ 4 \pi  N_\mathrm{Sk} F_e &- I_0 B j \Big( \cos{\gamma} ( 4 \pi  N_\mathrm{Sk} s_{\theta} + D_0 \alpha s_{r} )\\
\label{eq:general_thieleeq_angular}
    &-\sin{\gamma} ( 4 \pi  N_\mathrm{Sk} s_{r} - D_0 \alpha s_{\theta} )\Big)\Big],
\end{align}
where $\Gamma = \frac{1}{b ( (4 \pi  N_\mathrm{Sk})^2 + (D_{0} \alpha)^2 }$.

Both Néel and Bloch skyrmions experience a radial and tangential motion depending on the orientation of the spin polarization that is ultimately determined by the current orientation. The skyrmions start to move on a spiral trajectory, and upon approaching the edges, if the applied current is small enough, the skyrmion will not be destroyed. In this case, the edge force increases until the radial force is compensated by

\begin{align}
\nonumber
    F_e = - \frac{I_0 B j}{D_0 \alpha} \Big[ &\cos{\gamma}( 4 \pi  N_\mathrm{Sk} s_r - D_0 \alpha s_{\theta})\\
\label{eq:edge_compensation}
    + &\sin{\gamma}(4 \pi  N_\mathrm{Sk}s_{\theta} + D_0 \alpha s_r )\Big],
\end{align}
with the skyrmions reaching an equilibrium radial coordinate $r_0$ when $\dot{r} = 0 $. At this position, they can either become static or continue along a circular motion along the disk edge, given by the angular velocity 

\begin{align}
\label{eq:theta}
    \dot{\theta} = - \frac{I_0 B j}{b D_0 \alpha r_0}( \cos{\gamma} s_r + \sin{\gamma} s_{\theta}),
\end{align}
obtained by replacaing Eq.~\eqref{eq:edge_compensation} into Eq.~\eqref{eq:general_thieleeq_angular}.


\blue{Equation~\eqref{eq:theta} reveals which spin polarization and helicity lead to a circular motion of a skyrmion along the edges. Note that the sign of the current determines whether the circular motion occurs at the inner or outer edge of the disk. These results are significant because they allow for a qualitative description of the circular dynamics without the need to characterize the complicated edge potential. Still, the edges are very important as they are needed to compensate the transverse motion caused by the topological charge of the skyrmion.} 

\blue{Lastly, we want to comment on an acceleration effect at the edges~\cite{yoo2017current} that causes a change in the skyrmion velocity due to a slight deformation of the skyrmion when it is pushed against the edges. We note that this acceleration is very brief and does not affect the results given by Eq.~\eqref{eq:theta}. Once the skyrmion reaches the radius $r_0$, the external force $F_e$ given by Eq.~\eqref{eq:theta} fully compensates the transverse motion, and there is no acceleration. This is also valid for the results discussed in Sec.~\ref{subsection:STT}.}

\subsection{Confirmation by micromagnetic simulations}
\label{subsection:micromagnetic_simulations}

\blue{Next, we proceed to analyze this analytical result based on micromagnetic simulations that have been} conducted using the GPU-accelerated software package Mumax3~\cite{vansteenkiste2011mumax,vansteenkiste2014design}, which solves the Landau-Lifshitz-Gilbert (LLG) equation
\begin{align}
\nonumber
\partial_t \mathbf{m}_i = &-\gamma_e \mathbf{m}_i \times \mathbf{B}^{i}_\mathrm{eff} + \alpha  \mathbf{m}_i \times  \partial_t \mathbf{m}_i\\
&+ \gamma_e \epsilon \beta[( \mathbf{m}_i \times \mathbf{s}) \times \mathbf{m}_i].
\label{eq:LLG_mumax3}
\end{align}
Here $\mathbf{B}^{i}_\mathrm{eff} = \delta F/M_s \delta \mathbf{m}_i$ is the effective field derived from the system's total free energy density $F$, given as the sum of exchange interaction, perpendicular magnetic anisotropy, the demagnetization energy, Dzyaloshinskii–Moriya interaction and the Zeeman energy. The parameters in Eq.~\ref{eq:LLG_mumax3} are: the gyromagnetic ration $\gamma_e = 1.760 \times 10^{11} $ T$^{-1}$s$^{-1}$ and $\epsilon \beta= \frac{\hbar \Theta_{SH}}{2 e d_z M_s}$; where $d_z$ is the thickness of the magnetic layer, $e$ the electron's charge, $\hbar$ Planck's constant, $M_s$ the saturation magnetization and $j_s = \Theta_\mathrm{SH} j$ the spin current with spin orientation $\mathbf{s}$ generated by the spin-Hall angle $\Theta_\mathrm{SH}$. The last term accounts for the current-induced spin-orbit torque. \blue{Note that the field-like torque term has been neglected because it is typically small and only has a significant effect once the textures deform during propagation~\cite{litzius2017skyrmion}. In our simulations, the skyrmion profile remains mostly rigid.}

The simulated Corbino disk is characterized by an inner disk radius of $r_a=40\,\mathrm{nm}$ and an outer disk radius of $r_b=200\,\mathrm{nm}$. The thickness is $d_z=1\,\mathrm{nm}$ for Néel skyrmions and $d_z=3\,\mathrm{nm}$ for Bloch skyrmions. The cell size is $1\,\mathrm{nm} \times 1\,\mathrm{nm} \times 1\,\mathrm{nm}$ in both instances. In our simulations, the coordinate system is centered, with its origin $\bm{r} = (0, 0)$, precisely at the geometric center of the Corbino disk.

In nature, Bloch skyrmions can appear in systems with broken inversion symmetry, like MnSi, stabilized by the bulk-type Dzyaloshinskii–Moriya interaction (DMI), determined by $E_\mathrm{DMI}^{b}=D_b\int  [\bm{m}(\bm{r}) \cdot ( \nabla \cross \bm{m}(\bm{r}))  ]\mathrm{d}^3r$, with $D_b=1.12$ mJ/m$^2$, or in centrosymmetric materials, where the interactions present are the exchange energy $E_\mathrm{ex}=A\int |\nabla \bm{m}(\bm{r})|^2 \mathrm{d}^3r$, the magnetocrystalline anisotropy $E_\mathrm{anis} =K_\mathrm{u} \int [1 -  m_{z}(\bm{r})^2 
]\mathrm{d}^3r$ and demagnetization fields. By contrast, Néel skyrmions are stabilized at interfaces, such as Co/Pt, where the broken inversion symmetry gives rise to an interfacial-type DMI, defined by $E_\mathrm{DMI}^{\mathrm{i}}=D_\mathrm{i}\int  [m_z(\bm{r}) \nabla \cdot \bm{m}(\bm{r}) - (\bm{m}(\bm{r}) \cdot \nabla) m_z(\bm{r})]\mathrm{d}^3r$. 

We simulate a stable Néel skyrmion at an interface of Co/Pt with magnetic parameters similar to Ref.~\cite{sampaio2013nucleation}, with $M_s = 0.58$ MA/m. The interactions present are the dipole-dipole interaction, an interfacial DMI $D_\mathrm{i}= -3.5 $ mJ/m$^2$, exchange constant $A_\mathrm{ex}= 15$ pJ/m and perpendicular magnetic anisotropy $K_\mathrm{u}= 0.8$ MJ/m$^3$. Conversely, we simulate Bloch skyrmions in centrosymmetric materials without DMI interaction. We use magnetic parameters similar to Ref.~\cite{gobel2019forming}, where $M_s = 1.4$ MA/m. The interactions are the exchange constant $A_\mathrm{ex}= 15$ pJ/m, perpendicular magnetic anisotropy $K_\mathrm{u}= 1.2$ MJ/m$^3$ and an external magnetic field in the negative z-direction $B_z = -50$ mT. %

In the following, two scenarios are examined: 
1. Radial current inducing tangential spin polarization. This is the technologically relevant scenario as the current can easily be applied from the center to the outer edge. 
2. Tangential current inducing radial spin polarization.

\subsubsection{Radial current causing tangential spin polarization} 
In the first scenario (Fig.~\ref{fig:spin_tangential}a), the skyrmions start at $\bm{r} = (- 85 \, \mathrm{nm}, - 85 \, \mathrm{nm})$  with current $j = \pm 2.0 \, \mathrm{GA}/\mathrm{m}^2$, where the sign $\pm$  refers to Néel and Bloch, respectively. The different signs in the current are chosen such that both skyrmions move towards the outer radius of the disk. Néel skyrmions experience a radial SOT force (green arrow, Fig.~\ref{fig:spin_tangential}a) and exhibit a motion towards the edges due to the SkHE. Eq.\eqref{eq:theta}, suggests that when the spin polarization is purely tangential, the Néel skyrmion gets stuck at the edges: $\dot{\theta}\rightarrow 0$. Our micromagnetic simulations verify that a Néel skyrmion eventually comes to a halt (Fig.\ref{fig:spin_tangential}b). This occurs as the edge force in the negative radial direction counterbalances the SOT force. Conversely, Bloch skyrmions experience a tangential SOT force (red arrow, Fig.~\ref{fig:spin_tangential}a). This leads to sustained motion along the edges with $\dot{\theta} \neq 0$ in Eq.~\eqref{eq:theta}. This behavior is confirmed in the micromagnetic simulations (Fig.~\ref{fig:spin_tangential}c).

\begin{figure*}[t!]
  \centering
    \includegraphics[width=\textwidth]{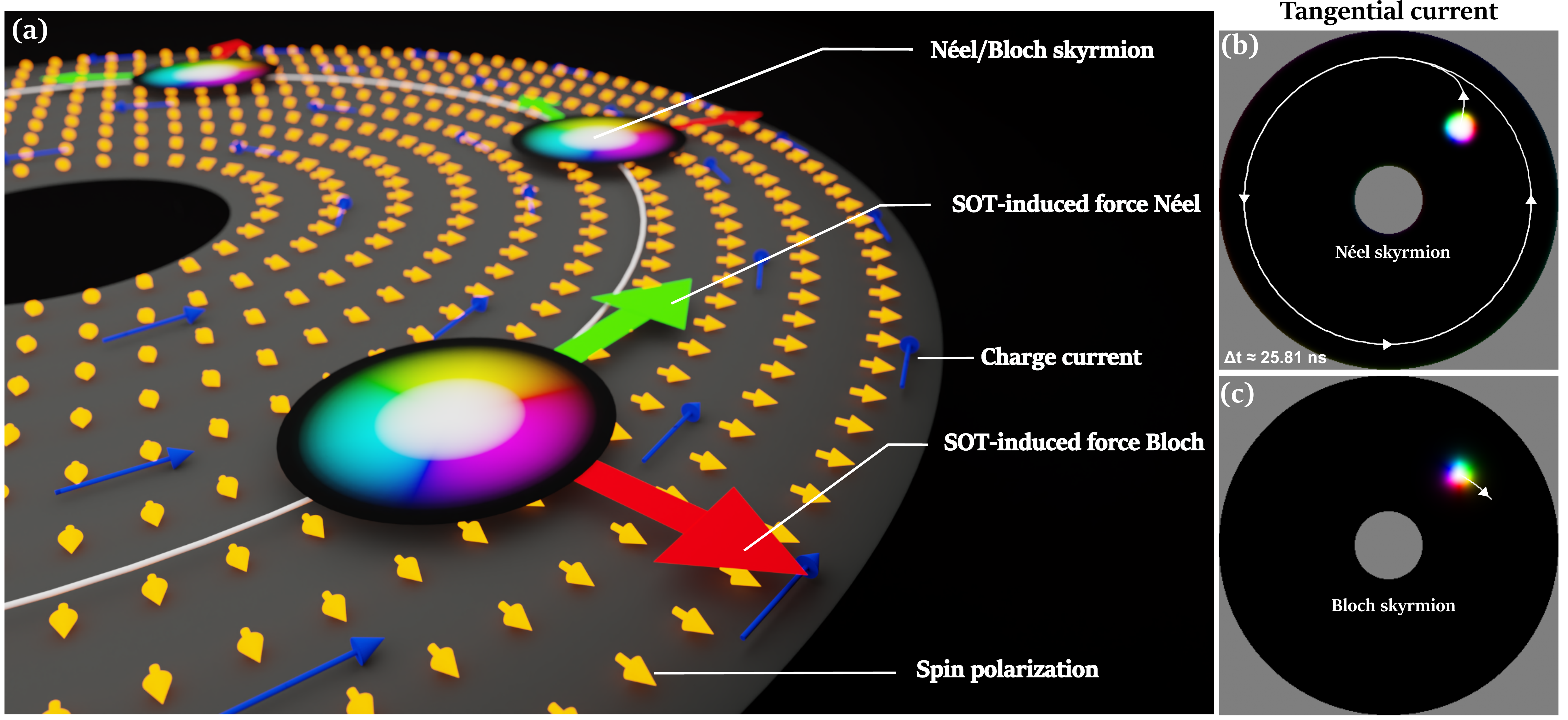}
    \caption{Dynamics of Néel and Bloch skyrmions under tangential currents generating a radial spin polarization. (a) Schematic representation of Néel and Bloch skyrmions subjected to spin currents with radial polarization. Néel skyrmions experience a tangential SOT force (green arrow), while Bloch skyrmions are influenced by a radial SOT force (red arrow). (b) and (c) show the trajectories of Néel and Bloch skyrmions, respectively, obtained by micromagnetic simulations for 100 ns for (b) and 60 ns for (c). The marker at the lower left corner of subplot (b) indicates the time duration for completing one loop around the disk with $\Delta t \approx 25.81$ ns for the Néel skyrmion.}
    \label{fig:spin_radial}
\end{figure*}

\subsubsection{Tangential current causing radial spin polarization} 
In the second scenario (Fig.~\ref{fig:spin_radial}a), the skyrmions start at $\bm{r} = (85 \, \mathrm{nm}, 85 \, \mathrm{nm})$, both with current $j = -2.0 \, \mathrm{GA}/\mathrm{m}^2$. In contrast to the previous case, Néel skyrmions experience a tangential SOT force (green arrow, Fig.~\ref{fig:spin_radial}a). They move toward the edges and continue to rotate due to $\dot{\theta} \neq 0$ in Eq.~\eqref{eq:theta}, as depicted in Fig.~\ref{fig:spin_radial} (b). On the other hand, Bloch skyrmions experience a radial SOT force (red arrow, Fig.~\ref{fig:spin_radial}a). The edge force counteracts this force, resulting in $\dot{\theta} = 0$ in Eq.~\eqref{eq:theta}. As a result, it is the Bloch skyrmion that gets stuck in this scenario. This dynamics is confirmed by our micromagnetic simulations illustrated in Fig.~\ref{fig:spin_radial} (c).

In our simulations, under the same current strength and with the edge force compensated, Néel skyrmions had a much higher frequency than Bloch skyrmions. Specifically, Néel skyrmions complete a full rotation under tangential current in  $\Delta t \approx 25.81$ ns, translating to a rotation frequency of $\approx 38.75$ MHz. In contrast, under a radial current, Bloch skyrmions take about $\Delta t \approx 132.1$ ns for a full rotation, corresponding to a frequency of  $\approx 7.6$ MHz. This indicates that the frequency of rotation for Néel skyrmions is about 5 times higher than that of Bloch Skyrmions for this set of parameters. However, the frequency of both skyrmions can be tuned by changing the magnetic parameters, since from Eq.~\eqref{eq:theta}, $\dot{\theta}$ is proportional to the ratio $I_0/D_0$. These quantities are determined by the skyrmion size. Thus, changing the magnetic interactions would lead to a change in the ratio and, consequently, to the frequency. Note that from Eq.~\eqref{eq:theta}, if the ratio $I_0/D_0$ and size of both types of skyrmions in Fig.~\ref{fig:spin_tangential} (c) and Fig.~\ref{fig:spin_radial} (b) were the same, they would have the same frequency.

In summary, our micromagnetic simulations are in agreement with the analytical predictions given by Eqs.~\eqref{eq:theta}. The torque tensor's shape, depending on the helicity [Eq.~\eqref{eq:torque_tensor_helicity}], dictates that the SOT-induced force relative to the spin polarization is (i) antiparallel/parallel for Bloch skyrmions and (ii) perpendicular for Néel skyrmions. Essentially, skyrmions move when the SOT-induced force is oriented perpendicularly to the edge force, requiring a radial current with tangential spin polarization for Bloch skyrmions and a tangential current with radial spin polarization for Néel skyrmions. Otherwise, the skyrmions get stuck at the edges. Due to the fact that a current can be applied radially much more easily than tangentially, the Bloch skyrmion rotating in a Corbino disk under radial current is the technologically more significant scenario identified by our calculations.

\subsection{Discussion of the circular skyrmion dynamics for radial currents  in Corbino disk devices}


Before we proceed with the discussion of skyrmioniums and the dynamics driven by spin transfer torque, we want to \blue{comment on a couple of points} that are \blue{important} to fully understand the skyrmion dynamics driven by SOT: 1. Circular skyrmion motion under realistic current densities, 2. The behavior of skyrmions with intermediate helicities between Bloch and Néel type, \blue{3. The comparison of the micromagnetic simulations with the Thiele equation, and 4.} The comparison of our findings with the skyrmion dynamics in conventional racetrack geometries.

\subsubsection{Radial currents in realistic systems} 
For the sake of simplicity, we have previously discussed constant current densities in the Corbino disk. However, especially in the technologically relevant case of a radial current, the current density is not constant but decreases with the distance as $r^{-1}$ due to the linearly increasing circumference with $r$. This effect can easily be accommodated in Eq.~\eqref{eq:Thiele} by replacing the current density $j$ with $j_a r_a/r$ where $r_a$ is the inner disk radius and $j_a$ the corresponding current density. This slightly alters the end result for $\dot{\theta}$ in Eq.~\eqref{eq:theta} as follows
\begin{align}
    \dot{\theta} = -\Big(\frac{j_a r_a}{r_0} \Big) \Big[\frac{I_0 B }{b D_0 \alpha r_0}( \cos{\gamma} s_r + \sin{\gamma} s_{\theta}) \Big].
\end{align}

 Therefore, the qualitative results discussed so far remain the same: While the Bloch skyrmion achieves a circular motion at $r_0$, the Néel skyrmion gets stuck. 

\subsubsection{Intermediate helicities between Bloch and Néel-type skyrmions} 
We have shown that in the case of the radial current, Bloch skyrmions achieve rotational dynamics. However, in reality, the interface of a centrosymmetric FM and the HM will give rise to an interfacial DMI such that the simulated Bloch-type skyrmions will not be perfectly Bloch-type. Instead, they are skyrmions with an intermediate helicity between Bloch and Néel type \cite{kim2018asymmetric}. In other words, depending on the strength of the interfacial DMI, the helicity could be somewhere between 0 and $\pi/2$ (or $-\pi/2$), where $\pi/2$ corresponds to the Bloch skyrmion simulated previously and 0 to a perfect Néel skyrmion. 

Our analytical results easily accommodate these variations. 
A radial current with tangential spin polarization with the same strength as before rotates the intermediate skyrmion around the disk's outer edge with a lower frequency. In this scenario, following from Eq.~\eqref{eq:theta}, as $\gamma$ decreases from $\pi/2$ (perfect Bloch skyrmion), so does $\dot{\theta}$. A decrease in $\gamma$ to 0 signifies a transition to a Néel skyrmion, at which point $\dot{\theta} $ decreases to 0. The maximum skyrmion rotational frequency in a Corbino disk under a radial current is achieved for a perfect Bloch skyrmion.

\subsubsection{\blue{Qualitative comparison with the Thiele equation}}

\blue{In Sec.~\ref{subsection:Corbino}, we have derived the Eq.~\eqref{eq:theta} for the angular velocity $\dot{\theta}$ analytically. In this section, so far, we have solely shown the skyrmion trajectory obtained via micromagnetic simulations. Circular trajectories have been achieved which means that Eq.~\eqref{eq:theta} has been validated qualitatively.} 

\blue{A quantitative comparison is also possible but brings about several numerical challenges, as explained below. Via micromagnetic simulations, we have obtained a frequency of approximately $38.75$ MHz for Néel skyrmion under tangential currents presented in Fig.~\ref{fig:spin_radial}(a). To obtain the frequency via Eq.~\eqref{eq:theta} based on the Thiele equation, one must determine the tensor coefficients $D_0$ and $I_0$ defined in Sec.~\ref{subsection:thiele_equation}. Estimating these quantities is problematic when the skyrmion is close to the edges as it is not clear where the non-collinear texture of the skyrmion ends and where the non-collinear texture at the edges starts. The integration required to compute $D_0$ and $I_0$ is impacted significantly by this choice. We obtain $D_0 \approx 22.49$ and $I_0 \approx 142.91$ nm that gives rise to a circular motion with the radius $r_0 \approx 169.3$ nm at a frequency of $39.77$ MHz, in very good agreement with the micromagnetic simulations.}

\blue{Equation~\eqref{eq:theta} is only valid for $\dot{r}=0$ when the gyroscopic force is compensated by the edge force. To fully compare the skyrmion trajectory with micromagnetic simulations, we determine the edge force in Eq.\eqref{eq:Thiele}, with $F_e = -\nabla E_\mathrm{total}$. The total energy is the sum of all interactions presented in the system. This term is a function of the skyrmion radial position, and it is calculated by fitting the total energy $E_\mathrm{total}$ against the skyrmion position. In our simulations, in the case of Fig.~\ref{fig:spin_radial}(a), we have determined this fit to be a sixth-order polynomial. Here again, the fit can be very sensitive to the selected data points and, to the best of our knowledge, there is no systematic method to identify the significant points.}

\blue{Overall, we are able to achieve excellent agreement between micomagnetic simulations and the semi-analytical Thiele equation. However, the numerical calculation of the coefficients $D_0$ and $I_0$ as well as the interaction potential with the edge is technical and not straight forward. Therefore, in the following we will focus on the results of micromagnetic simulations that can be understood qualitatively by Eq.~\eqref{eq:theta} based on the Thiele equation.}

\subsubsection{Comparison with rectangular racetrack geometries}

\blue{The antagonistic dynamics of Néel and Bloch skyrmion in rectangular samples, like in a racetrack storage device, has already been analyzed in Ref.~\cite{tomasello2014strategy}. In the following, we review these findings in the context of our results and compare them. In the racetrack memory, the applied current is homogeneous and ideally flows along its length. In such a device, one aims to move the skyrmion along the current direction, perpendicular to $\bm{s}$. In contrast, within a Corbino disk, in the more realistic scenario of a radial current, one aims to move the skyrmion perpendicular to the current direction in a circular motion along $\bm{s}$.}


Interestingly, our results suggest that these geometries require different skyrmions. In Corbino disks, the Bloch skyrmion moves along the edges, and the Néel skyrmion gets stuck (Fig.~\ref{fig:spin_tangential}). Conversely, in a racetrack, it is the Néel skyrmion that can propagate along the edge of a racetrack \cite{tomasello2014strategy,gobel2019overcoming} while the Bloch skyrmion gets stuck \cite{de2023biskyrmion}. This is because, for Néel skyrmions, the SOT force is oriented along the current direction and perpendicular to the spin polarization. On the other hand, for Bloch skyrmions, the SOT force is perpendicular to the current direction and parallel to the spin polarization. Therefore, in a racetrack, the SOT force is pointing along the desired direction of motion (along the track) for the Néel skyrmion, and in the Corbino disk geometry, the SOT force points along the desired direction (tangentially) for the Bloch skyrmion. In both cases, due to the SkHE, the skyrmions will move towards the edges at the beginning, where the edge force increases until the transverse force component due to the topological charge is compensated.

Lastly, we would like to close this discussion by commenting on the more unconventional scenario in which the current is applied tangentially. In a Corbino disk, Bloch skyrmions get stuck at the edge, and Néel skyrmions engage in the circular motion (Fig.~\ref{fig:spin_radial}). For comparison, we consider a racetrack geometry where the current is applied along its width rather than its length. This causes $\bm{s}$ to be parallel to the racetrack's length direction. Under these conditions, it is the Néel skyrmion that gets stuck. In this scenario, only a Bloch skyrmion could propagate along the edges because now, the SOT force points along the desired direction of motion --along the track.

Consequently, the optimal strategy for manipulating the skyrmion motion at the edges depends on the specific geometry and current orientation. To facilitate the motion of Néel skyrmions along the edges, the current should be oriented tangentially for Corbino disks and along its length for racetracks. However, in turn, this causes Bloch skyrmions to get stuck. Conversely, to ensure the motion of Bloch skyrmions, the current should be aligned radially for Corbino disks and along the width of racetracks. These scenarios lead to Néel skyrmions getting stuck.

\section{Beyond skyrmions: skyrmioniums}
\label{subsection:skyrmioniums}

\begin{figure}[h!]
  \centering
    \includegraphics[width=\columnwidth]{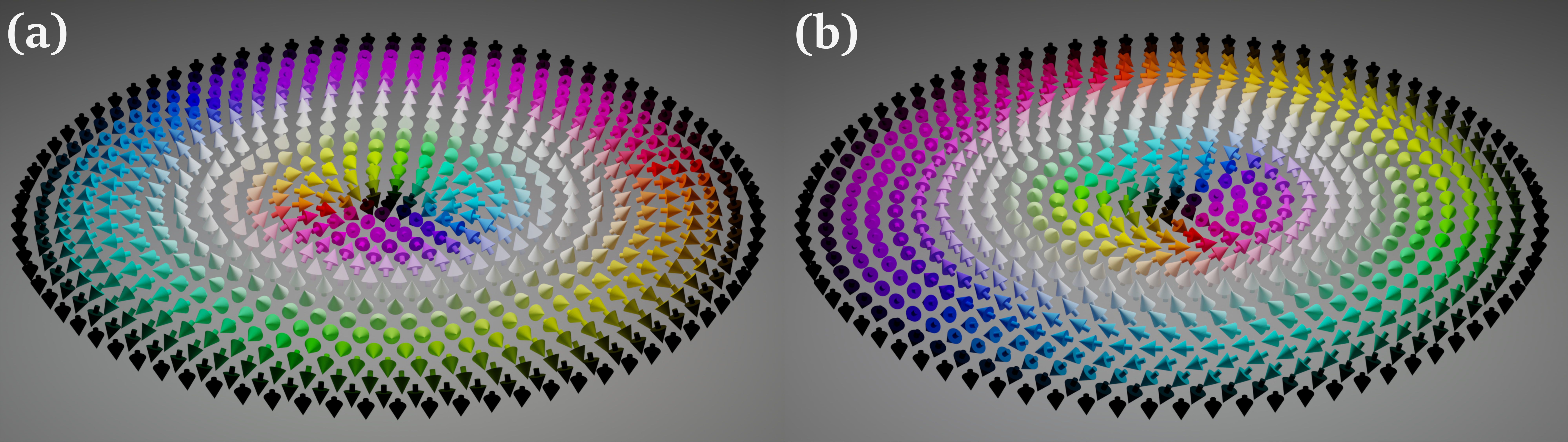}
    \caption{Illustrative comparison of two types of magnetic skyrmioniums. Schematic representation of (a) Néel-type skyrmionium and (b) Bloch-type skyrmionium. Arrows correspond to the orientation of magnetic moments like in Fig.~\ref{fig1:skyrmions}.}
    \label{fig:skyrmioniums}
\end{figure}

\begin{figure}[t]
  \centering
    \includegraphics[width=\columnwidth]{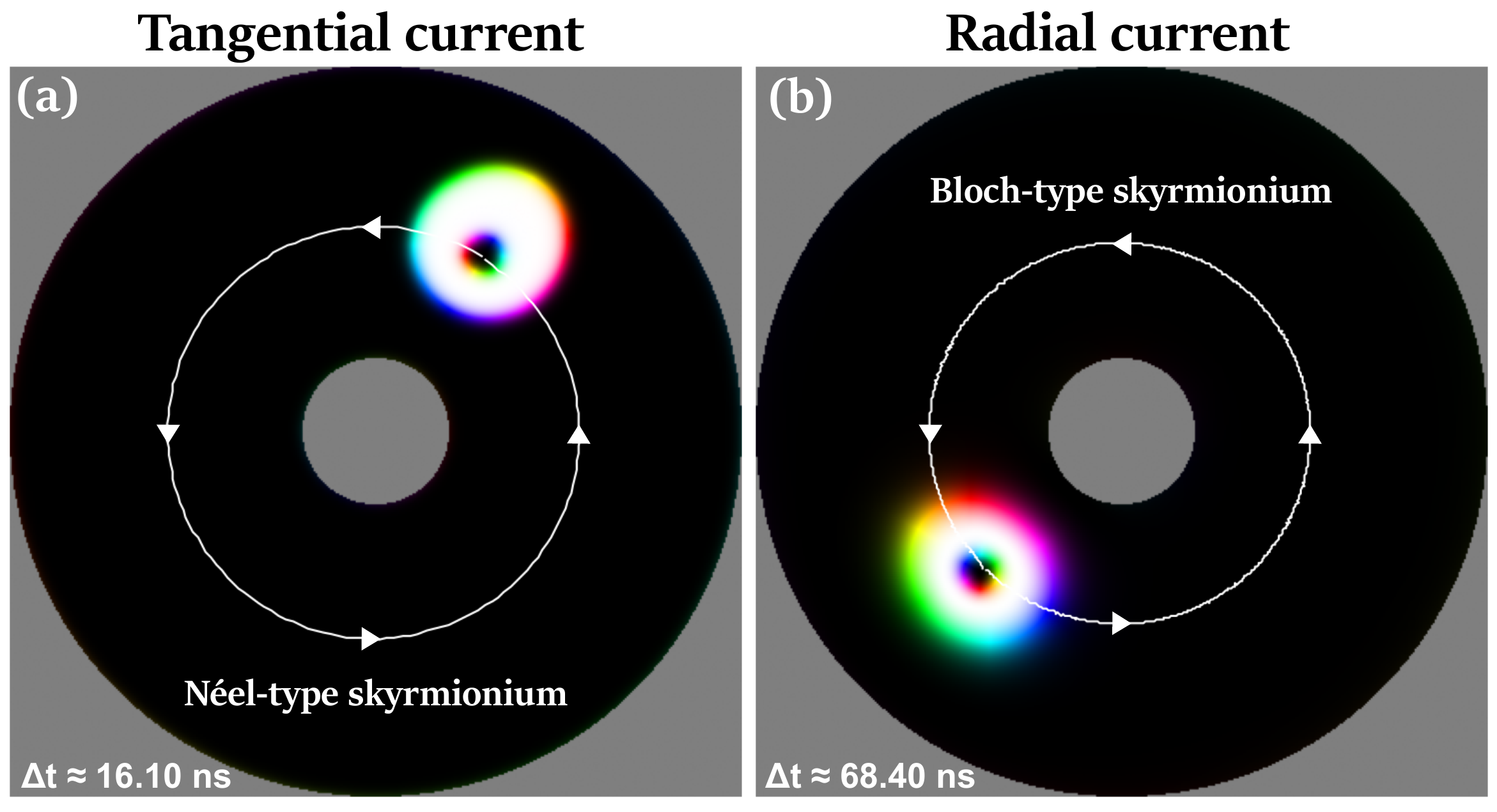}
    \caption{Dynamics of skyrmionium in Corbino disks influenced by spin-orbit torque (SOT). The figure illustrates the trajectory for one full rotation of (a) a Néel-type skyrmionium subjected to tangential current (radial spin polarization) and (b) a Bloch-type skyrmionium under the influence of a radial current (tangential spin polarization). The markers at the lower left corner of each subplot indicate the time duration for completing one loop around the disk: $\Delta t \approx 16.10$ ns for Néel-type (a) and $\Delta t \approx 68.40$ ns for Bloch-type (b). The difference in $\Delta t$ in (a) and (b) arises from distinct stabilizing interaction parameters, impacting the skyrmioniums' size and ratio $I_0/D_0$ in Eq.~\eqref{eq:theta_skyrmionium}.}
    \label{fig:skyrmionium_simulations}
\end{figure}

In this section, we want to address the significance of the skyrmion's topology for the SOT-induced motion. While a skyrmion has a topological charge $N_{\text{sk}}= \pm 1$, a skyrmionium is topologically trivial $N_{\text{sk}}=0$ \cite{gobel2019electrical}. A skyrmionium consists of a skyrmion in the center and a skyrmionic ring at the outside. Therefore, overall, the topological charge is compensated, but the magnetic texture is still very stable. In the following, we will explore how the motion of skyrmioniums differs from the motion of skyrmions. Fig.~\ref{fig:skyrmioniums} schematically represents a skyrmionium, which can either be of Néel or Bloch type, depending on the underlying magnetic material.

The vanishing topological charge of skyrmioniums makes the gyroscopic vector in Eq.~\eqref{eq:Thiele} disappear, thereby rendering the skyrmion Hall effect zero. As a result, these magnetic quasi-particles move in the direction of the force that is applied due to the SOT. This force is quantified by the torque tensor $\underline{\bm{I}}$. Skyrmioniums are similar to skyrmions except that they exhibit a $2\pi$ rotation of the azimuthal angle between the center and their edge—in contrast to the $\pi$ rotation of standard skyrmions. Thus, the torque tensor has the exact symmetry as in Eq.\eqref{eq:torque_tensor_helicity}, except that $I_0$ is larger. Therefore, the SOT force for Néel- and Bloch-type skyrmioniums points in the same direction as their skyrmionic counterparts presented in Figs.~\ref{fig:spin_tangential} and ~\ref{fig:spin_radial}. 

This allows us to understand the motion of both types of skyrmioniums in Corbino disk geometries. We start from the Thiele equation in polar coordinates and disregard the edge force since it only results in a short radial motion toward the point where it vanishes. For a general spin polarization, the radial velocity yields
\begin{align}
\label{eq:radial_skyrmionium}
    \dot{r} =  \frac{I_0 B j }{b D_0 \alpha} ( \cos{\gamma} s_{\theta} -  \sin{\gamma} s_{r}),   
\end{align}
and the angular velocity $\dot{\theta}$
\begin{align}
\label{eq:theta_skyrmionium}
    \dot{\theta} = -\frac{I_0 B j}{r b  D_0 \alpha } (\cos{\gamma} s_{r} + \sin{\gamma} s_{\theta}).
\end{align}
Note that Eq.~\eqref{eq:theta_skyrmionium} is the same as Eq.~\eqref{eq:theta}, indicating that the skyrmionium dynamics is qualitatively the same as the skyrmion motion presented in the previous section. Similar to skyrmions, a radial current with a tangential spin polarization enables a circular motion only for Bloch-type skyrmioniums. Conversely, a Néel-type skyrmionium moves along the radial direction toward the edge of the disk, where it gets stuck. The opposite dynamics occur for tangential currents with radial spin polarization: Néel-type skyrmioniums engage in the circular motion, while a Bloch skyrmionium gets stuck at the edge.

These results are confirmed by micromagnetic simulations presented in Fig.~\ref{fig:skyrmionium_simulations}. Here, we simulate Néel and Bloch-type skyrmioniums within the framework discussed in the previous section, employing identical initial conditions, parameters, and external currents. The only parameter adjustment that was necessary to stabilize a Bloch skyrmionium in Fig.~\ref{fig:skyrmioniums}(b) is the incorporation of a bulk DMI and an increase of external field $B_z=-80$ mT. In these simulations, circular motion is unique to Néel-type skyrmioniums under a tangential current with radial spin polarization, as demonstrated in Fig.~\ref{fig:skyrmionium_simulations}(a). In contrast, when the current is radial, i.\,e., the spin polarization is tangential, this motion is exclusive to Bloch-type skyrmioniums, as depicted in Fig.~\ref{fig:skyrmionium_simulations}(b).

\begin{figure*}[t!]
  \centering
    \includegraphics[width=\textwidth]{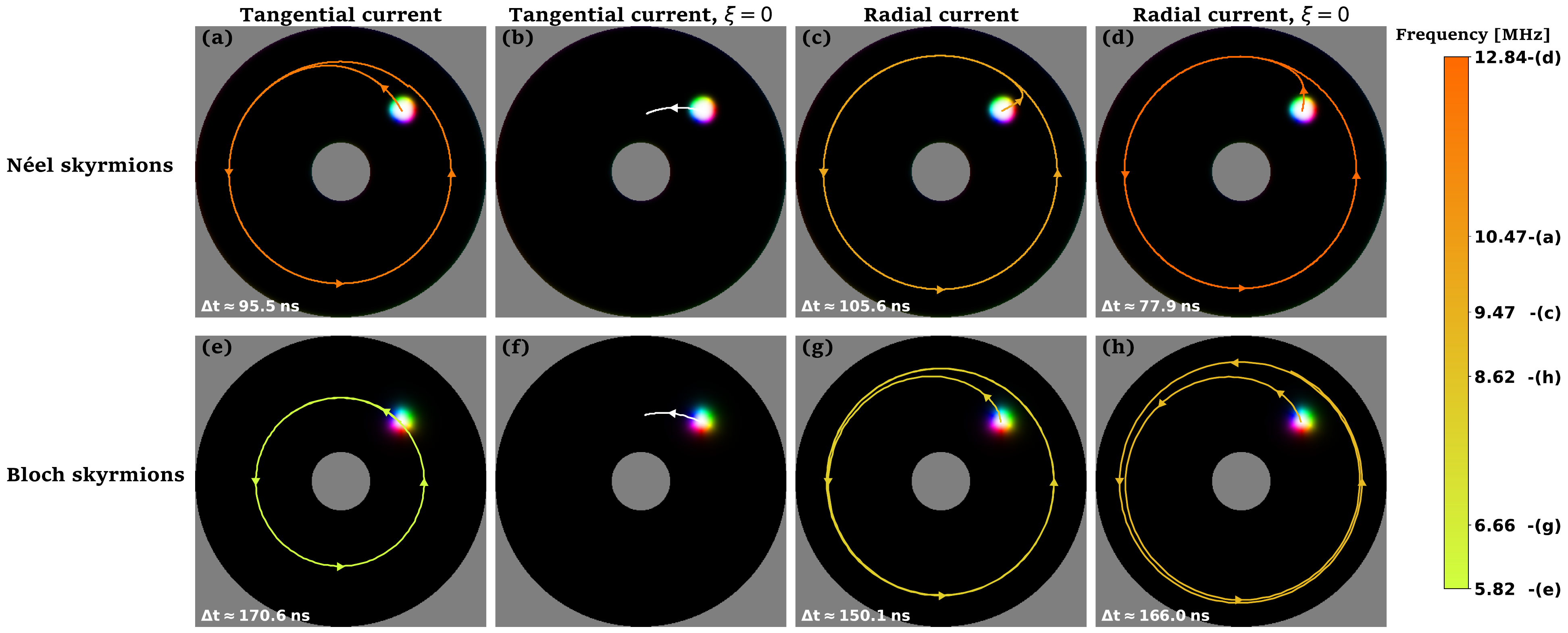}
    \caption{Dynamics of Néel and Bloch skyrmions in Corbino disks under STT with radial and tangential currents. (a-d) Show the trajectory of a Néel skyrmion under (a) a tangential current, (b) a tangential current with non-adiabaticity constant $\xi =0$, (c) a radial current, and (d) a radial current with $\xi =0$. The panels from (e-h) display the same for Bloch skyrmions. The trajectory's color corresponds to the rotation frequency, as denoted by the color bar. Markers at the lower left corner of each subplot indicate the time duration required for completing one loop around the disk. The fastest scenario is observed in (d), while the slowest is in (e). As both types of skyrmions become immobilized at the inner edges, markers are omitted in (b) and (f), and the trajectory is plotted in white after a simulation time of $30$ ns for (b) and $46$ ns for (f).}
    \label{fig:STT}
\end{figure*}

While the dynamics of skyrmioniums and skyrmions are similar, magnetic skyrmioniums are very attractive since we have obtained Eq.~\eqref{eq:theta_skyrmionium} without the need for edge repulsion. This is advantageous, especially when considering higher current densities. Regular skyrmions might get destroyed if the applied current density is too large, as the edge force may no longer be sufficient to compensate for the SOT force. Consequently, this allows higher current values to be used with skyrmioniums without risking destruction. Additionally, in our simulations, the frequency of rotation of skyrmioniums is higher than for conventional skyrmions, even for the same current density. For instance, consider the scenario with a tangential current for the Néel skyrmion in Fig.\ref{fig:spin_radial}(b). In this case, the frequency is $\approx 38.75$ MHz. In contrast, a Néel-type skyrmionium, as in Fig.\ref{fig:skyrmioniums}(a), under the same parameters and current strength, exhibits a higher frequency of $\approx 62.11$ MHz. This is because although Eqs.~\eqref{eq:theta} and \eqref{eq:theta_skyrmionium} are the same, the ratio $I_0/D_0$ changes for skyrmioniums due to the larger size compared to skyrmions. Moreover, besides tuning the frequency by changing the magnetic parameters, one could also tune the frequency by changing the initial radius position of the skyrmionium, since from \eqref{eq:theta_skyrmionium} the frequency is inversely proportional to $r$.

\section{Current-induced motion by spin transfer torques (STTs)}
\label{subsection:STT}

In our previous discussions of SOT-driven skyrmions and skyrmioniums, helicity played a pivotal role in determining the motion in Corbino disk geometries. We now turn our attention to an equally compelling scenario: the dynamics of skyrmions driven by spin transfer torque (STT) in similar geometries, with currents applied in both radial and tangential directions. The STT case is known to be less efficient compared to the SOT scenario due to a mostly adiabatic propagation of the spin-polarized currents~\cite{sampaio2013nucleation}. Still, the STT scenario is technologically important because it does not require an additional HM layer, which might make it easier to realize for an experimental realization. Here, the current flows directly in the FM and gets spin-polarized, as schematically illustrated in Fig.~\ref{fig:schematic_scenarios} (b). Since the magnetic texture is non-collinear, this spin-polarized current generates a torque that allows it to move skyrmions. Another advantage is that Bloch skyrmions might be easier to realize in the STT scenario because the interface required in the SOT scenario typically gives rise to interfacial DMI, favoring Néel skyrmions, which get stuck under SOT. 

In contrast to the SOT scenario, Néel and Bloch skyrmions exhibit similar dynamics under STT. This similarity can be attributed to the absence of a helicity-dependent term in the Thiele equation given by
\begin{equation}\label{eq:Thiele_STT}
    b \underline{\bm{G}} \times (\bm{u} - \bm{v}) + b \underline{\bm{D}} (\xi  \bm{u} - \alpha \bm{v})   = \bm{F}_e.
\end{equation}
Here, $\bm{u}$ is the velocity of the conduction electrons, defined as $\bm{u} = b_J \bm{j}$. The constant $b_J$ is defined via $b_J = \mu_B P / e M_s (1+\xi^2)$, where $\mu_B$ is the Bohr-magneton, $P$ the current's spin polarization and $\xi$ is the non-adiabaticity parameter of the STT~\cite{zhang2004roles}. Similarly to the previous section, we consider the motion driven by radial and tangential currents. Thus, it is convenient to express this equation in polar coordinates
\begin{align}
  -4 \pi  N_\mathrm{Sk} b
&\begin{pmatrix}
 u_{\theta} - r  \dot{\theta} \\
\dot{r} - u_{r}
\end{pmatrix}
+ b D_{0} 
\begin{pmatrix}
\xi u_{r}  -\alpha  \dot{r}\\
 \xi u_{\theta} - \alpha r  \dot{\theta}
\end{pmatrix}
=
 \begin{pmatrix}
 F_e\\
0
\end{pmatrix},
\end{align}
where $u_r$ and $u_{\theta}$ are the radial and tangential velocities of the conduction electrons, respectively. The solution for the radial velocity yields
\begin{align}
\nonumber
    \dot{r} = \Gamma \Big[ - D_0 \alpha F_e &+ b ( (4 \pi  N_\mathrm{Sk})^2 + D^2_0 \alpha \xi)u_r  
    \\&+ b 4 \pi  N_\mathrm{Sk} D_0(\xi -\alpha)u_{\theta}   \Big],
    \label{eq:STT_radial}
\end{align}
and the angular velocity $\dot{\theta}$
\begin{align}
\nonumber
    \dot{\theta} = \frac{\Gamma}{r} \Big[ 4 \pi  N_\mathrm{Sk} F_e &+ ((4 \pi  N_\mathrm{Sk})^2 + D^2_0 \alpha \xi) u_{\theta} 
    \\&+4 \pi  N_\mathrm{Sk} D_0(\alpha -\xi) u_r  \Big],
\end{align}
where $\Gamma = \frac{1}{b ( (4 \pi  N_\mathrm{Sk})^2 + (D_{0} \alpha)^2 }$.

From the equations above, in the absence of the edges, one can deduce that Néel and Bloch skyrmion will have a spiral trajectory as analyzed in Ref.~\cite{kechrakos2023skyrmions}. However after some time, the edge force compensates the transverse force due to the topological charge, stopping the motion in the radial direction. From Eq.~\eqref{eq:STT_radial} we find in that case
\begin{align}
\nonumber
    F_e = \frac{b}{D_0 \alpha} \Big[ &D_0 4 \pi N_\mathrm{Sk}(\xi - \alpha) u_{\theta}  \\
\label{eq:edge_compensation_stt}
     &+((4 \pi  N_\mathrm{Sk})^2 + D^2_0 \alpha \xi) u_r\Big],
\end{align}
hence, the angular velocity at the equilibrium position is given by
\begin{align}
\label{eq:theta_STT}
    \dot{\theta} =  \frac{1}{D_0 \alpha r_0}(4 \pi N_\mathrm{Sk} u_r + D_0 \xi u_{\theta} ).
\end{align}

While skyrmions can move along a circular trajectory in the STT scenario as well as in the SOT scenario, a notable difference arises: under STT, the rotational motion is independent of the helicity, allowing Néel and Bloch skyrmions to move along the edges. This is true for radial as well as tangential currents. Still, it is worth highlighting a critical contrast for these current types in the STT case. Therefore, we simulate the behavior of Néel and Bloch skyrmions under STT within a Corbino disk geometry with the same parameters as before in Sec.~\ref{subsection:micromagnetic_simulations}. The 
simulations are conducted as in Sec.~\ref{subsection:micromagnetic_simulations} but for a modified Landau-Lifshitz-Gilbert (LLG) equation given by
\begin{align}
\nonumber
\partial_t \mathbf{m}_i = &-\gamma_e \mathbf{m}_i \times \mathbf{B}^{i}_\mathrm{eff} + \alpha  \mathbf{m}_i \times  \partial_t \mathbf{m}_i
\\\nonumber
&-\frac{b_J}{M_s}[\mathbf{m}_i \times (\mathbf{m}_i \times (\mathbf{j} \cdot \mathbf{\nabla}) \mathbf{m}_i)]
\\
\label{eq:LLG_mumax3_STT}
&-\xi b_J[\mathbf{m}_i \times (\mathbf{j} \cdot \mathbf{\nabla}) \mathbf{m}_i],
\end{align}
where the last two terms account for the current-induced spin transfer torque. 

The skyrmion motion is much slower, so we increase the current to $j=-20.0$ GA/m$^2$, 10 times larger than in the SOT scenario, with polarization 0.35~\cite{li2004domain} and the non-adiabaticity constant set to $\xi=0.6$.

In the case of a radial current, the angular velocity is $\dot{\theta} = 4 \pi N_\mathrm{Sk} u_r/(D_0 \alpha r_0)$, indicating that the motion is primarily governed by the skyrmions' intrinsic properties, including topological charge $N_\mathrm{Sk}$ and the skyrmion shape, quantified by dissipation tensor component $D_0$. Importantly, circular motion is exclusive to topologically non-trivial spin textures characterized by a non-zero topological charge ($N_\mathrm{Sk} \neq 0$). Yet, at the same time, the topological charge induces the transverse motion with respect to the applied force, a fundamental aspect of the skyrmion Hall effect. Interestingly, it is this underlying mechanism that allows the skyrmions to keep moving along the edges. Consequently, Néel and Bloch skyrmions exhibit a circular motion, confirmed by the results of micromagnetic simulations presented in Figs.~\ref{fig:STT}(c) and (g), respectively.

For a tangential current, the angular velocity is $\dot{\theta} = u_{\theta} \xi/( \alpha r_0)$. The non-adiabaticity torque parameter, $\xi$, primarily influences the circular motion. The micromagnetic simulations in Figs.~\ref{fig:STT}(a) and (e) illustrate this relationship for Néel and Bloch skyrmions, respectively. Both skyrmion types can achieve circular motion as long as $\xi \neq 0$. On the other hand, if $\xi = 0$, both skyrmions come to a halt as they encounter the edges as illustrated in Figs.~\ref{fig:STT}(b) and (f).

Besides this qualitative sinificance of the non-adiabacity constant $\xi$, the constant also enters Eq.~\eqref{eq:theta_STT} via $u_r = \mu_B P j_r/ e M_s (1+\xi^2) $. While setting $\xi =0$ halts the skyrmions at the edges for a tangential current, a radial current not only allows a circular motion irrespective of $\xi$ but also benefits from a small non-adiabacity. As depicted in Figs.~\ref{fig:STT}(d) and (h), when $\xi = 0$, our micromagnetic simulations show that the frequency for Néel skyrmions is $\approx 12.84$ MHz [Fig.~\ref{fig:STT}(d)]. In comparison, with $\xi = 0.6$, the frequency drops to $\approx 9.47$ MHz [Fig.~\ref{fig:STT}(c)]. From Eq.~\eqref{eq:theta_STT} one can calculate that the velocity increases as $\sim (1 + \xi^2)$.

Lastly, we would like to comment on Néel- and Bloch-type skyrmionium dynamics under STT. As discussed previously, these magnetic textures move directly along the direction of the applied force due to the absence of a topological charge. They are pushed along the direction of the conduction electrons current $\bm{u}$. Accounting for the vanishing topological charge, Eq.~\eqref{eq:Thiele_STT} leads to $\dot{\theta} = (\xi/ r_0 \alpha) u_{\theta} $ and $\dot{r} = (\xi/\alpha) u_{r}$. Therefore, we draw two conclusions: (i) under radial current, both skyrmionium types will not be able to move since such current only moves them towards the edges, and (ii) they can only undergo a circular motion under tangential currents, as long as $\xi \neq 0$. Interestingly, in the STT scenario, skyrmioniums do not behave exactly as their skyrmion counterparts. Moreover, in the case of a radial current, the absent topological charge is detrimental to achieving a circular motion. As discussed previously, skyrmions can keep moving along the edges due to gyroscopic force in Eq.~\eqref{eq:Thiele_STT}, while skyrmioniums get stuck.

In summary, the STT case also offers the possibility for a circular motion, however, with fundamental differences compared to the SOT scenario. In the technologically relevant case of radial currents, both skyrmion types move along the edge of the disk due to their topological charge. Topologically trivial skyrmioniums, on the other hand, do not allow for this type of motion for radial currents due to the absence of the skyrmion Hall effect.

\begin{table*}[t!]
  \centering
  \caption{Summary of the discussed scenarios. $\xmark$ means the skyrmion(ium) does not engage in a circular motion. The columns correspond to the type of the applied charge current (tangential or radial), the type of torque-driven motion, and, in the case of the spin transfer torque, the non-adiabaticity parameter $\xi$. The numerical values are the results of our micromagnetic simulations. The frequencies cannot be compared one to one because the skyrmions and skyrmioniums have different sizes and shapes.}
\setlength\tabcolsep{6.5pt} 
\small 
\begin{tabularx}{\textwidth}{|X|c|c|c|c|c|c|}
  \hline
   \textbf{Skyrmion-type}& {\textbf{ \thead{Radial\\ SOT}}} & {\textbf{ \thead{Tangential \\ SOT}}} & {\textbf{ \thead{Radial  \\ STT [$\xi=0.6$]}}} & {\textbf{ \thead{Tangential \\ STT [$\xi=0.6$]}}} & {\textbf{ \thead{Radial  \\ STT [$\xi=0$]}}} & {\textbf{ \thead{Tangential  \\ STT [$\xi=0$]}}} \\
  \hline
  Néel skyrmion     & \xmark & \thead{38.75 MHz \\ Fig.~\ref{fig:spin_radial} (b)} & \thead{9.47 MHz \\ Fig.~\ref{fig:STT} (c)} & \thead{10.47 MHz \\ Fig.~\ref{fig:STT} (a)} & \thead{12.84 MHz \\ Fig.~\ref{fig:STT} (d)} & \xmark \\
  \hline
  Bloch skyrmion    & \thead{7.6 MHz \\ Fig.~\ref{fig:spin_tangential} (c)} & \xmark & \thead{6.66 MHz \\ Fig.~\ref{fig:STT} (g)} & \thead{5.82 MHz \\ Fig.~\ref{fig:STT} (e)} & \thead{8.62 MHz \\ Fig.~\ref{fig:STT} (h)} & \xmark \\
  \hline
  Néel-skyrmionium  & \xmark & \thead{62.11 MHz \\ Fig.~\ref{fig:skyrmionium_simulations} (a)} &  \xmark & \thead{14.28 MHz } & \xmark & \xmark \\
  \hline
  Bloch-skyrmionium & \thead{14.62 MHz \\ Fig.~\ref{fig:skyrmionium_simulations} (b)} & \xmark & \xmark & \thead{6.28 MHz } & \xmark & \xmark \\
  \hline
\end{tabularx}
\label{tab:summary}
\end{table*}

\section{Conclusions}\label{sec:conclusion}

In summary, we have analyzed the motion of skyrmions and skyrmioniums in Corbino disks analytically and based on micromagnetic simulations. In Sec.~\ref{section:SOTskyrmion}, we analyzed Bloch and Néel skyrmions driven by SOT. Néel skyrmions are known to be able to creep along the sample's edge in a rectangular sample, making them more suitable for racetrack memories than Bloch skyrmions, which get stuck. However, as we have shown, Bloch skyrmions are more relevant in Corbino disks. Under the application of a radial current, i.\,e., tangential spin polarization, they carry out a circular motion along the edge of the nano-disk. Néel skyrmions, on the other hand get stuck at the edge. Yet, when the spin polarization points along the radial direction, e.g., by applying a tangential current, the roles are reversed: In this case, the Bloch skyrmions get stuck at the edge, and the Néel skyrmions carry out a circular motion. 

In Sec.~\ref{subsection:skyrmioniums}, we have generalized these results to the topologically trivial skyrmioniums. These non-collinear spin textures behave similarly to their skyrmion counterparts when driven by SOT. However, due to the absence of a topological charge, a Bloch-type skyrmionium under radial current will move along a circle without being pushed towards the edge. The absence of the SkHE is helpful in this scenario as it allows an increase in the current density and rotation frequency without causing the skyrmionium to annihilate at the edge.

In Sec.~\ref{subsection:STT}, we have considered skyrmions and skyrmioniums driven by STT. Here, the spin polarization follows the magnetic texture instead of being always perpendicular to the applied current. Again, we have observed circular motion for skyrmions, but this time, Néel and Bloch skyrmions behave similarly. Only in the synthetic case of a tangential current, without the non-adiabaticity of the spin polarization(i.e., $\xi=0$), will the skyrmions get stuck at the edge. In all realistic scenarios, it is noteworthy that the SkHE, often viewed as detrimental, plays a crucial role in enabling skyrmions to move along the disk's edge. This finding has been confirmed by the fact that skyrmioniums will always get stuck when driven by STT caused by radial currents. This unexpected utility of the SkHE in facilitating circular motion marks an important insight into the significance of topology, underscoring the potential of leveraging this effect in spintronic applications. 

An overview of all the discussed scenarios of this paper is shown in Tab.~\ref{tab:summary}. We expect the SOT scenario of radial currents to be most relevant technologically \blue{for spin torque nano-oscillators}, especially using skyrmioniums, because lower current densities are required to make the non-collinear object move along a circular trajectory with an MHz frequency.

\section*{Acknowledgements}
This work is supported by SFB TRR 227 of Deutsche Forschungsgemeinschaft (DFG) and SPEAR ITN.
This project has received funding from the European Union’s Horizon 2020 research and innovation program under the Marie Skłodowska-Curie grant agreement No 955671. 
\blue{This work was supported by the EIC Pathfinder OPEN grant 101129641 ``OBELIX''.}
Membership in the International Max Planck Research School for Science and Technology of Nano-Systems is gratefully acknowledged. I.A. performed the simulations. B.G. and I.A. wrote the manuscript with significant input from all authors. I.A. prepared the figures. All authors discussed the results. B.G. and I.M. planned and supervised the project.

\bibliography{references.bib} 
\bibliographystyle{naturemag}

\end{document}